# Silane-Catalyzed Fast Growth of Large Single-Crystalline Graphene on Hexagonal Boron Nitride

Shujie Tang,[1,2] Haomin Wang,[1,*] Hui Shan Wang,[1,3] Qiujuan Sun,[1,3] Xiuyun Zhang,[4] Chunxiao Cong,[5] Hong Xie,[1] Xiaoyu Liu,[1] Xiaohao Zhou,[6] Fuqiang Huang,[7] Xiaoshuang Chen,[6] Ting Yu,[5] Feng Ding,[4] Xiaoming Xie,[1,8,*] and Mianheng Jiang [1,8]

[1] State Key Laboratory of Functional Materials for Informatics, Shanghai Institute of Microsystem and Information Technology, Chinese Academy of Sciences, 865 Changning Road, Shanghai 200050, P.R. China

[2] Graduate University of the Chinese Academy of Sciences, Beijing 100049, P.R. China

[3] School of Physics and Electronics, Central South University, Changsha, 410083, P. R. China

[4] Institute of Textiles and Clothing, Hong Kong Polytechnic University, Kowloon, Hong Kong, 999077, P. R. China

[5] Division of Physics and Applied Physics, School of Physical and Mathematical Sciences, Nanyang Technological University, 21 Nanyang Link, Singapore 637371

[6] National Laboratory for Infrared Physics, Shanghai Institute of Technical Physics, Chinese Academy of Sciences, Shanghai 200083, P. R. China

[7] CAS Key Laboratory of Materials for Energy Conversion, Shanghai Institute of Ceramics, Chinese Academy of Sciences, Shanghai 200050, P.R. China

[8] School of Physical Science and Technology, ShanghaiTech University, 319 Yueyang Road, Shanghai 200031, P. R. China

*Electronic mail: hmwang@mail.sim.ac.cn, xmxie@mail.sim.ac.cn

**Abstract:** The direct growth of high-quality, large single-crystalline domains of graphene on a dielectric substrate is of vital importance for applications in electronics and optoelectronics. Traditionally, graphene domains grown on dielectrics are typically only ~1 µm with a growth rate of ~1 nm/min or less, the main reason is the lack of a catalyst. Here we show that silane, serving as a gaseous catalyst, is able to boost the graphene growth rate to ~1 µm/min, thereby promoting graphene domains up to 20 µm in size to be synthesized via chemical vapor deposition (CVD) on hexagonal boron nitride (*h*-BN). Hall measurements show that the mobility of the sample reaches 20,000 cm$^2$/V·s at room temperature, which is among the best for CVD-grown graphene. Combining the advantages of both catalytic CVD and the ultra-flat dielectric substrate, gaseous catalyst-assisted CVD paves the way for synthesizing high-quality graphene for device applications while avoiding the transfer process.

As the first isolated atomically thin crystal,[1] graphene has attracted enormous interest from people all over the world because of its rich physical properties and great potential in various applications. Among them, applications in electronics are the most appealing, though they are challenging because they require high-quality, large-area samples.[2] Among the popular methods of graphene synthesis, chemical vapor deposition (CVD) is known as the most promising for scalable growth of high-quality graphene sheets.[3–18] Progress in this area, as demonstrated by the successful synthesis of 30-inch continuous films,[19] the wafer-scale growth of a single-crystal monolayer of graphene on Ge,[20] and centimeter-sized graphene domains,[15] has been achieved. However, post-growth transfer remains a major hurdle for further applications because transfer processes introduce unavoidable surface/interface contamination, cracks, and excessive wrinkles in graphene. Moreover, owing to the rough catalytic metal surface, the quality of the graphene grown is still far from the flatness of the mechanically exfoliated samples. Thus, the ability to develop a transfer-free technique by growing graphene directly on dielectric substrates with a large domain size, fast growth rate and very high quality is highly desirable.

Recently, molybdenum disulfide, tungsten disulfide and *h*-BN are found to be good candidates for the substrate of graphene devices.[21] Among them, hexagonal boron nitride (*h*-BN) is obviously the best one because of its insulation properties, chemical inertness and small lattice misfit. Growing graphene directly on ultra-flat *h*-BN can greatly preserve the pristine properties of graphene, and the mobility of the aligned graphene grown on *h*-BN via the CVD method reaches 20,000–30,000 $cm^2/V·s$ at room temperature.[22] However, this growth method suffers from the great drawback of very low growth rate (~1 nm/minute) and the generation of domain sizes that are mostly smaller than 1 μm. While the fabrication of a graphene domain 11 μm in size has been achieved, it took 72 h to complete the fabrication.[23] Such a slow growth rate, which is three to four orders of magnitude slower than that obtained by CVD on a catalytic metal surface, originates mainly from the lack of a catalyst in the process.

In this report, we demonstrate that silane and germane as gaseous catalysts can boost the growth rate of graphene on *h*-BN by almost 2-order of magnitude higher than that in absence of gaseous catalyst. The introduction of the gaseous catalyst also effectively improves the percentage of precisely aligned domains on *h*-BN. These results present a promising route towards very high-quality, transfer-free graphene on dielectric substrates for electronic applications.

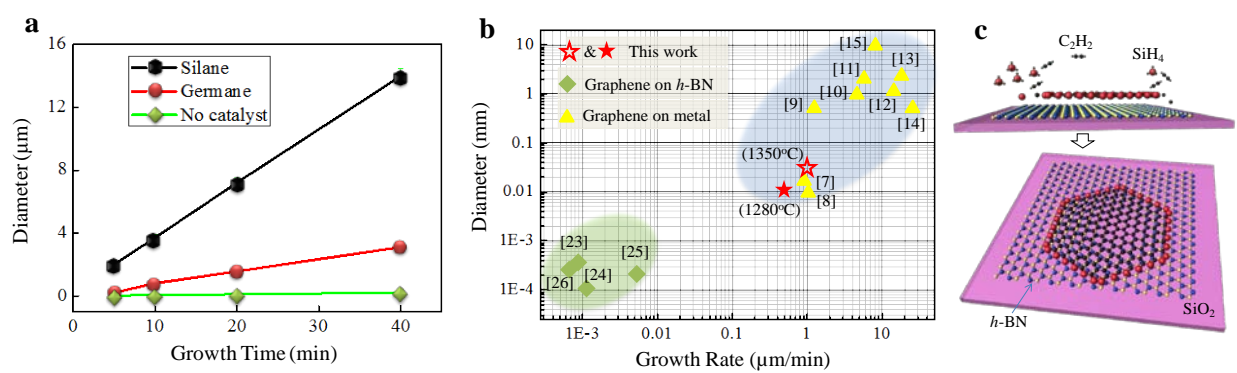

**Figure 1 | The gaseous catalyst-assisted CVD of graphene on *h*-BN.** (a) The growth duration dependence of the domain dimensions for single-crystalline graphene in the presence of silane (black) or germane (red) gaseous catalysts and no catalyst (green). The diameter of the graphene domain is measured as the diagonal length of the hexagonal graphene crystal, and the growth temperature is 1280°C. (b) The graphene growth rates plotted as a function of the dimensions of the single-crystalline domains obtained in this work and those reported in the literature. The data

points are labeled with the reference number from whence they came in brackets. (c) Schematic of the gaseous catalyst-assisted chemical vapor deposition process, visualizing the carbon (black), nitrogen (blue), boron (yellow), silicon (red), and hydrogen (small gray) atoms as spheres in the illustration. The schematic illustrates simplified schemes of the catalytic growth of monolayer graphene onto $h$-BN, where silicon atoms from the decomposition of $SiH_4$ attach to the edge of the graphene and assist its growth.

## Results

**Gaseous catalyst assisted graphene growth.** A series of experiments are carried out to understand the effects of different gaseous catalysts upon the growth of graphene. Fig. 1a presents the dependence of the domain size upon the growth time at a growth temperature of 1280°C. It is found that the domain size always increases linearly with growth time. Also, it is found that the growth rate of graphene in the absence of a catalyst is about 5 nm/min, which increases by an order of magnitude to 50 nm/min when germane is introduced. If silane is introduced, the growth rate further increases to 400 nm/min and, if the substrate is further heated to 1350°C, the growth rate reaches ~1000 nm/min, which is very close to that of graphene grown on a metal surface by CVD (Fig. 1b). At such a high growth rate, the largest domain size of the graphene grown on $h$-BN reaches 20 μm after 20 min of growth (Fig. 1b). As shown in Fig. 1b,[7-14,24–27] using silane as a catalyst improves both the graphene growth rate and the typical domain size on $h$-BN by two orders of magnitude when compared with no catalyst. Auger electron spectroscopy is used to evaluate the amount of Si and Ge residue in the grown graphene samples (see Supplementary Fig. 4), whereupon no Si or Ge signal is detected, confirming that silane and germane are suitable to serve as catalysts for graphene growth. It is worthwhile to note that excessive supply of silane will cause formation of SiC nanoparticles (see Supplementary Fig. 3). The GCA-CVD method thereby bridges the huge gap between the direct growth of graphene on a dielectric substrate and that on a metal surface, and the graphene growth on $h$-BN presented here has already entered into the practical application regime. Fig. 1c schematically illustrates the mechanism of graphene growth on $h$-BN, where the silicon atoms attach to the edge of the graphene domain and serve as the catalyst to reduce the reaction barrier for $C_2H_2$ molecules to form the honeycomb lattice along the graphene edge during growth.

**Simulation of growth mechanism.** To achieve a more detailed understanding of the role of silicon in graphene CVD growth, density functional theory calculations are performed on armchair graphene, with the results shown in Fig. 2. Without a catalytic metal surface, the graphene growth front must be passivated by H atoms as the reaction is carried out in a hydrogen-rich environment. The chosen carbon feedstock, $C_2H_2$, ensures the formation of a complete hexagon for each cycle of $C_2H_2$ incorporation at the graphene edge. Our calculation indicates that three reaction steps are required to incorporate a $C_2H_2$ molecule onto the graphene edge to form a new hexagon, with corresponding energy barriers for these reaction steps of 5.80, 3.17 and 5.80 eV. In the case of a silicon catalyst (Fig. 2b), when a silicon atom attaches to the growth front, only two steps are required to incorporate a $C_2H_2$ onto the graphene edge to form a new hexagon with corresponding energy barriers of 2.68 and 3.38 eV. Because each reaction step is exothermic, the threshold barriers for $C_2H_2$ incorporation onto a graphene edge without and with a Si atom as a catalyst are 5.80 and 3.38 eV, respectively. These calculations clearly indicate that this much-reduced threshold barrier when a catalyst is involved is responsible for the increased growth rate. Graphene edge activation by metal ad-atoms has been studied previously by simulations.[28–31] Experimentally, copper vapor was used in the demonstration of graphene growth on insulating substrates.[32,33] Oxygen was another surface activator, accelerating the graphene growth on Cu and suppressing its nucleation at the same time.[15] The use of silane/germane, conventional gases used in semiconductor industry with good accessibility and controllability, has not been previously proposed.

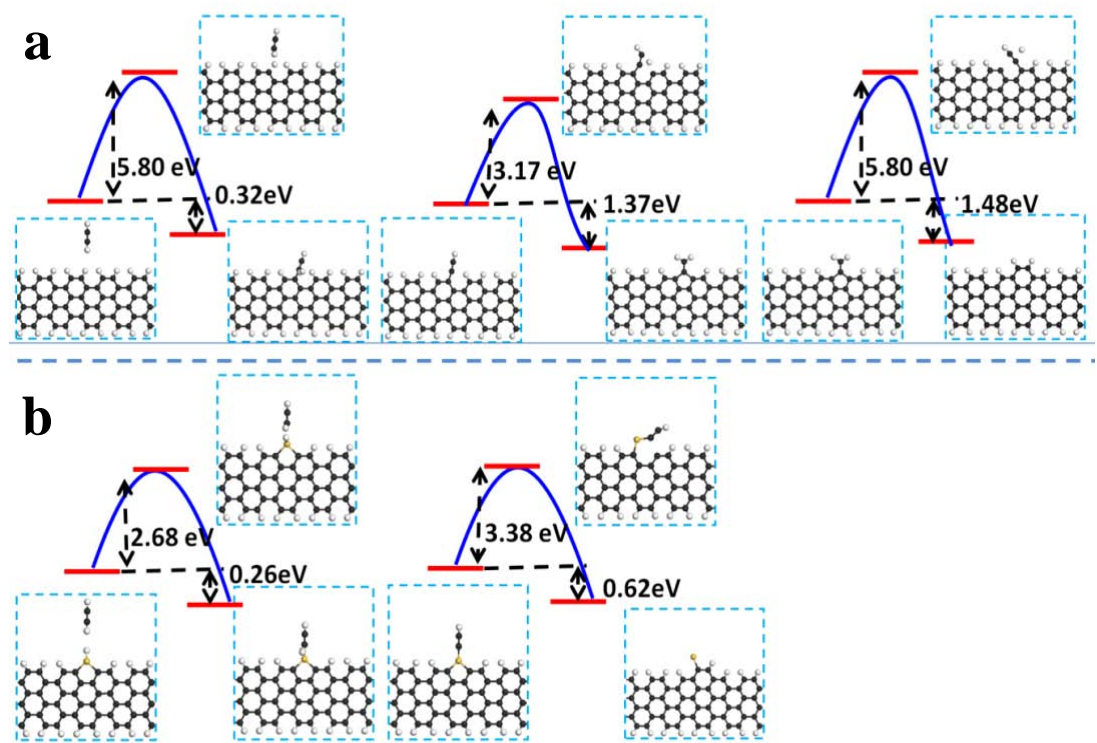

**Figure 2 | Density functional calculation of the reaction between a $C_2H_2$ molecule with the commonly observed armchair graphene edge** (a) without and (b) with a Si atom as a catalyst on the graphene edge. The black, white and yellow balls represent carbon, hydrogen and silicon atoms, respectively.

**Atomic force microscopy characterization and alignment survey.** The crystallinity and alignment of the graphene domains can be studied by analyzing the moiré patterns obtained by atomic force microscopy (AFM), where the integrity of the pattern gives information of graphene crystallinity and the alignment can be judged by the periodicity and the rotation of the moiré pattern with respect to the *h*-BN lattice.[22] M. Yankowitz *et al*. also did systematic investigations of the graphene/*h*-BN super-lattice with different angular rotation.[34] In our experiments, three types of domains are observed. The "A" domains present a regular hexagonal shape with a giant moiré pattern, whose edges are along the armchair direction. The periodicity of the moiré pattern is about 13.9 nm, which demonstrates that these graphene domains are precisely aligned with the underlying *h*-BN.[22] The "B" domains also exhibit a regular hexagonal shape but with no detectable moiré patterns, and detailed atomic-resolution AFM measurements indicate that all of these "B" graphene domains are rotated ~30° with respect to the underlying *h*-BN lattice. The "C" domains exhibit the typical polycrystalline structure, with a detectable moiré pattern only on some sub-domains. Fig. 3a shows a typical "A" domain with a giant moiré pattern, shown in Fig. 3b, with the atomic-resolution image of the graphene and the underlying *h*-BN shown in Fig. 3c and 3d, respectively. The alignment of the moiré pattern with respect to the *h*-BN is very sensitive to the angular separation between the graphene and *h*-BN when the angular separation is less than 1°.[22] Although the measurement may possess a ±3° measurement error for the lattice orientation, mis-orientation between the graphene and *h*-BN should be less than 0.05° by evaluating the alignment of the moiré pattern with respect to the *h*-BN. These results prove that the graphene is precisely aligned with the underlying *h*-BN. Details about the other two types of domains are given in Supplementary Fig. 6, Supplementary Fig. 7 and Supplementary Discussion.

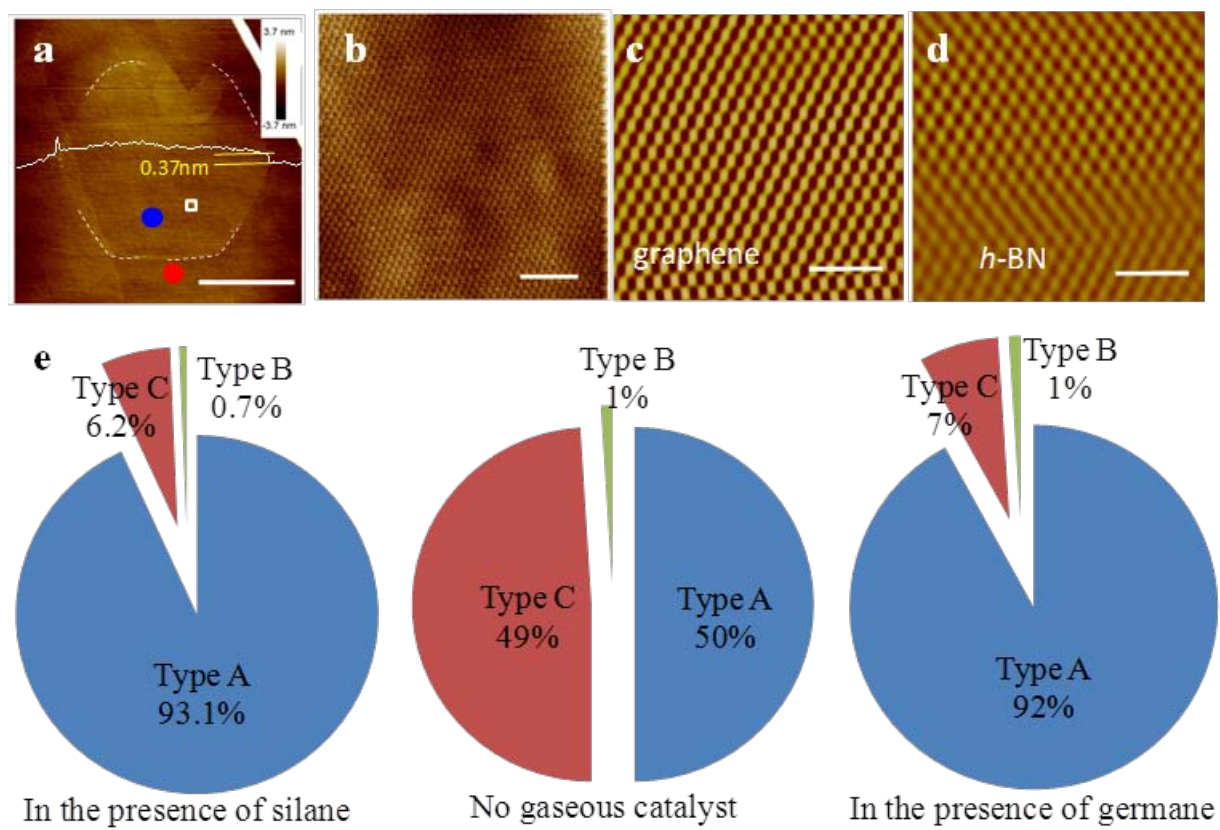

**Figure 3 | The investigation on crystallinity and alignment of graphene domains grown on *h*-BN.** (a) Topography image of a typical single-crystalline graphene domain with a diameter of 20 μm. The dashed line frames the shape of the graphene grain. There is a small second layer on the upright corner of the domain. The line scan shows the graphene thickness to be 0.37 nm. The scale bar is 10 µm. (b) The AFM friction image of the selected area in white box of (a), where the giant moiré pattern with a periodicity of 13.9 nm can be clearly seen. The scale bar is 100 nm. Atomic-resolution AFM images of (c) graphene and (d) *h*-BN taken from the areas marked by the blue and red dots in panel (a), respectively. During the measurement, the scanning angles are always kept the same. The scale bars are 1 nm. (e) Pie charts of the distribution of the type of

graphene domains obtained with/without gaseous catalysts. Type "A" indicates a graphene domain that is precisely aligned with the underlying *h*-BN, type "B" is one whose lattice is rotated ~30° with respect to the underlying *h*-BN lattice, and type "C" is one with a polycrystalline structure.

Fig. 3e shows the statistics of the types of graphene domains grown with and without gaseous catalysts. In the absence of a gaseous catalyst, the precisely aligned domains (Type A) and polycrystalline domains (Type C) appear in almost equal probability and only ~1% domains are the 30°-rotated single-crystalline domain (Type B). When using silane or germane as a catalyst, however, the percentage of precisely aligned domains (Type A) greatly improves to 93.1 or 92%.

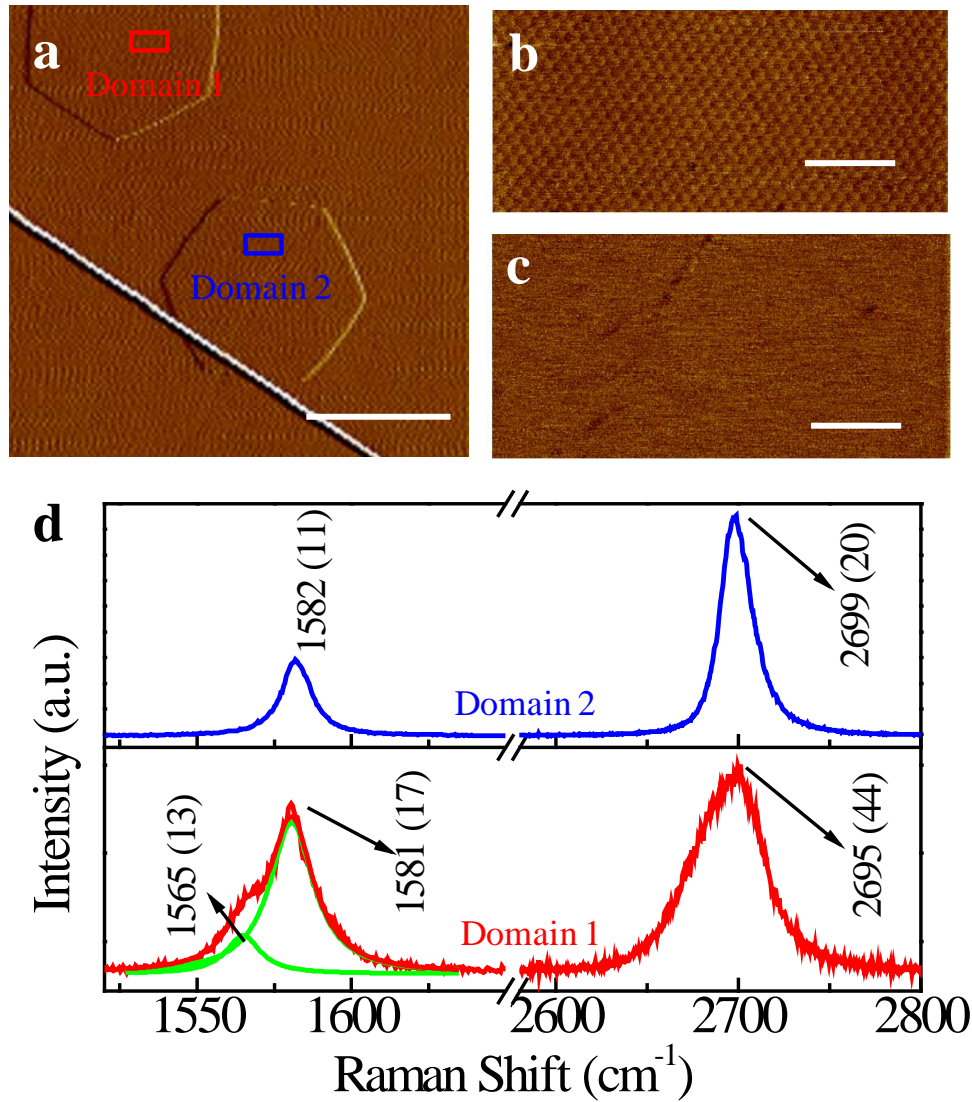

**Figure 4 | Raman analyses on graphene domains grown on *h*-BN.** (a) Topography images of two graphene domains grown on an *h*-BN surface. The white line across the image indicates a wrinkle on the *h*-BN surface formed during cooling. The scale bar is 2 µm. (b) Magnified view from the red box in (a), where the presence of a giant moiré pattern indicates precisely aligned graphene with respect to the *h*-BN. (c) Magnified view from the blue box in (a), where no detectable moiré pattern is seen. The scale bars in (b) and (c) are 100 nm. (d) Raman spectra taken from Domain 1 (lower plot) and Domain 2 (upper plot) in (a). The full width at half maximum for each peak is given in parentheses with the peak location value, and the wavelength of the exciting laser is 488 nm.

**Raman characterization of graphene domains grown on *h*-BN.** Fig. 4 shows the Raman spectra from Domain 1 (which is precisely aligned and noted as "type A") and Domain 2 (type "B", which is a 30°-rotated single-crystalline domain). The Raman results for the aligned or misaligned graphene domains grown on *h*-BN by the GCA-CVD method are very similar to those for aligned or misaligned graphene flakes on *h*-BN made by mechanical exfoliation,[35] indicating that the GCA-CVD-grown graphene domains are of very high quality. It can be observed that the Raman spectrum of precisely aligned graphene exhibits obvious differences from that of rotated domains. In the precisely aligned graphene, the full width at half maximum (FWHM) of the 2D band is ~44 cm$^{-1}$, which is broader than that of graphene domains with a large misalignment angle. Moreover, a shoulder peak located at 1565 cm$^{-1}$ that is non-dispersive (see Supplementary Fig. 8) is noticeable, which can be attributed to a transverse optical phonon and may be activated by folding.[35,36] In addition, a more comprehensive analysis of the Raman results of the GCA-CVD-grown graphene domains is given in Supplementary Fig. 8 and Supplementary Table 2.

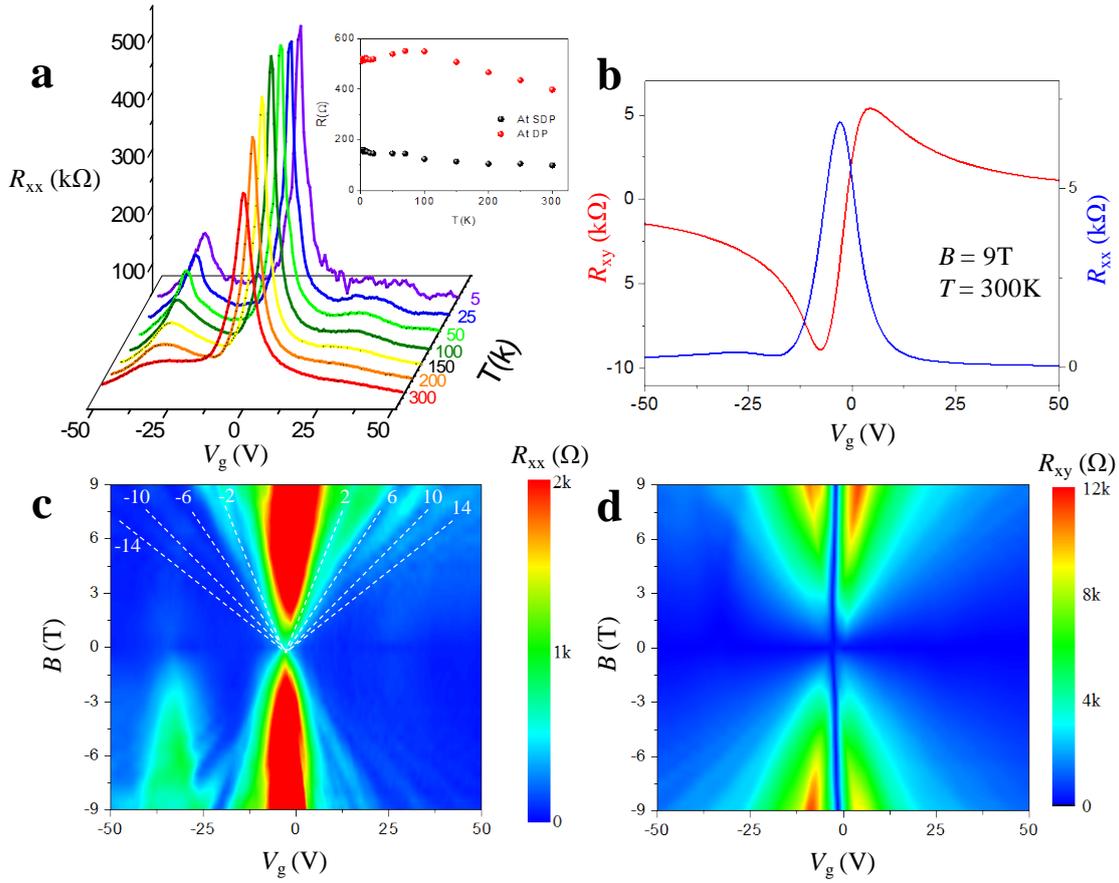

**Figure 5 | Transport measurement of the single-crystalline graphene precisely aligned with the underlying *h*-BN.** (a) Back-gate voltage ($V_g$) dependence of the longitudinal resistance at different temperatures. Inset shows the temperature dependence of the resistance at the Dirac point (DP) (red spheres) and satellite peaks at the hole doping (black spheres). (b) Longitudinal ($R_{xx}$, blue) and Hall resistance ($R_{xy}$, red) versus $V_g$ at temperature $T$ = 300 K and magnetic field $B$=9 T. (c–d) Quantum Hall effect fan diagram of (c) $R_{xx}$ and (d) $R_{xy}$ as a function of $V_g$ and $B$ at a temperature of 2 K.

**Electronic transport properties.** A Hall bar device is made on a heavily doped silicon substrate for measuring the Hall and field effect mobilities (for details of the device fabrication, see methods & Supplementary Fig. 9). Fig. 5 shows the field effect characteristics of the device (Fig. 5a), the results of the Hall measurements at a temperature of 300 K and a magnetic field of 9 T (Fig. 5b), and the effects of Landau level (LL) splitting upon the longitudinal and Hall resistances at a temperature of 2 K (Fig. 5c–5d). The carrier-independent field effect mobility can be calculated from the plot in Fig. 5a, and is found to be 17,000 cm$^2$V$^{-1}$s$^{-1}$ at 300 K. The hole and electron mobilities can be derived from Fig. 5b, with values of 19,000 and 23,000 cm$^2$V$^{-1}$s$^{-1}$, respectively. The high mobility value measured for these samples indicates that the electrical quality of the graphene domains on *h*-BN is among the best for CVD-grown graphene[15,37,38] and is comparable to that of micromechanically exfoliated graphene[39] (see Supplementary Table 3).

Two side peaks are observed in Fig. 5a owing to secondary Dirac cones, which are a result of the graphene/*h*-BN moiré pattern, and are consistent with other reports.[40–42] It is worth noting that the resistances at the Dirac Point (DP) do not significantly increase with decreasing temperatures. Woods *et al.* have reported similar results, which in their study were explained as a result of the suppression of the commensurate state.[43]

In the plots of $R_{xx}$ and $R_{xy}$ as a function of both gate voltage and magnetic field, the standard Quantum Hall effect (QHE) for graphene is observed to exhibit valleys in $R_{xx}$ (Fig. 5c) and plateaus in $R_{xy}$ (Fig. 5d) at the filling factors $v= \pm 4(n+1/2) = \pm 2, \pm 6, \pm 10\ldots$, where $n = 0, 1, 2\ldots$ is the LL index. A further investigation to demonstrate the effects of the secondary Dirac point upon the magnetic field is still underway.

**Conclusion**

In summary, by combining the advantages of the catalytic CVD method and an ultra-flat dielectric substrate, we have developed a novel method to synthesize graphene domain sizes up to 20 µm on *h*-BN with a growth rate comparable to that of graphene growth on metal surfaces and a graphene quality comparable to that made by the mechanical exfoliation method. In addition, the synthesized graphene on *h*-BN is transfer-free, which allows further electronic applications based on graphene/dielectric hetero-structures. Most of the graphene domains are perfectly aligned with the *h*-BN. No grain boundaries are observed when two domains merge; this brings hope to future fabrication of single crystalline graphene/h-BN wafer, which is essential for its scalable application on electronics.

**Methods**

**Graphene synthesis.** Before graphene growth, *h*-BN flakes were mechanically exfoliated onto quartz by Nitto tape and, after acetone cleaning, the quartz substrate with the *h*-BN flakes was loaded into the growth system. The graphene growth was carried out in a low-pressure chemical vapor deposition (LP-CVD) furnace at a temperature of 1100–1400°C. The system was then heated to 1280°C under an Ar/$H_2$ (5:1) mixture flow of 10 cm$^3$ per min (sccm), corresponding to 15 Pa, and annealed for 5 min, after which the Ar/$H_2$ flow was turned off. The $C_2H_2$ flow and a mixture of silane/argon or germane/argon (5% mole ratio of silane or germane to argon) were introduced into the system for the graphene growth. The pressure was kept at 5 Pa during the growth, and the growth time was in the range of 5–40 min. After growth, both the $C_2H_2$ and gaseous catalyst flow were turned off and the system was cooled down to room temperature with the Ar/$H_2$ mixture flowing. The samples of graphene on *h*-BN grown on a quartz surface were first transferred to a highly doped p-type silicon wafer with a 300-nm-thick $SiO_2$ capping layer for electrical transport studies.

**Scanning probe microscopy.** As-grown samples were characterized by an AFM (Dimension Icon, Bruker) in contact mode, and the atomic-resolution images were obtained by the AFM (Multimode III, Veeco) under ambient conditions. The AFM images on different surfaces were recorded in contact mode using SNL-10 AFM tips from Bruker that possess a nominal tip radius of less than 10 nm, and whose cantilevers possess a force constant $k = 0.05–0.5$ N/m. To obtain a high accuracy, scanners with a travel range less than 10 µm along the X and Y directions were used. Calibration for the atomic resolution was performed with newly cleaved highly ordered pyrolytic graphite (HOPG) prior to the measurements where, after calibration, the mean distance between the vicinal carbon atoms was measured to be 0.142 nm. The integral gain and set-point were adjusted to be as low as possible to obtain optimal images, and the scan rate was set to a value in the range of 10–60 Hz to reduce the noise from thermal drift. Several hours of pre-scanning were carried out to warm up the scanner and ensure high imaging stability. Error in this orientation measurement varies by data points and was typically ±2°, though it sometimes reached ±3°.

**Density functional theory calculation.** All of the calculations were performed within the framework of the density functional theory (DFT) as implemented in the Vienna Ab initio Simulation Package (VASP). The exchange-correlation potentials were treated by the gradient density approximation, and the interaction between valence electrons and ion cores was described by the projected augmented wave method. The energy cutoff for the plane wave functions was 400 eV and the force acting on each atom was less than 0.02 eV/Å. The energy barrier of every advance step was explored by climbing nudged elastic band method. The vacuum layer inside the super-cell was kept as large as 14 Å to avoid interaction with the adjacent unit cell. The Brillion zone was sampled as 1×2×1 grid meshes for the 7×5 slab and 1×1×1 grid meshes for the larger 8×6 and 7×8 slabs using the Monkhorst-Pack scheme during the calculation.

**Raman spectroscopy and mapping.** Raman spectra were obtained with a WITec micro-Raman instrument possessing excitation laser lines of 488/532/633 nm. An objective lens of 100× magnification and a 0.95 numerical aperture (NA) was used, producing a laser spot that was ~0.5 µm in diameter. The laser power was kept less than 1 mW on the sample surface to avoid laser-induced heating. The Raman images were acquired using a WITec Raman system with a 600 lines/mm grating and a piezo crystal-controlled scanning stage under a 488-nm laser excitation. The scanning step was about 200 nm.

**Device fabrication and electronic transport measurements.**
The Hall bar structure of the graphene devices was defined by a standard electron beam lithographic technique. The metal contacts (60 nm Au/10 nm Ti) were deposited through electron beam evaporation, whereupon the devices were annealed in a hydrogen atmosphere at 250°C for 3 h to remove resist residues and to reduce the contact resistance between the graphene and electrodes prior to the electrical measurements. Because the thickness of the $h$-BN flake on the 300-nm-thick SiO$_2$/Si substrates is about 20 nm, the effective capacitance, C$_g$, can be estimated as 10.5 nF•cm$^{-2}$. Both the electrical transport and magneto-transport measurements were carried out in a physical property measurement system (PPMS from Quantum Design, Inc.) via the standard lock-in technique.

**Acknowledgments:** We thank Y.B. Zhang from Fudan University for fruitful discussions. The work at the Shanghai Institute of Microsystem and Information Technology, Chinese Academy of Sciences, is partially supported by the National Science and Technology Major Projects of China (Grant No. 2011ZX02707), the Chinese Academy of Sciences (Grant Nos. KGZD-EW-303 and XDB04040300), and the projects from the Science and Technology Commission of Shanghai

Municipality (Grant Nos. 12JC1410100 and 12JC1403900).

**Author contributions:** M.J. and X.X. directed the research work. H.W., X.X. and S.T. conceived and designed the experiments. S.T. fabricated the graphene samples. S.T. and H.W. performed the AFM experiments. H.W., H.X. and X.L. fabricated the electronic devices. H.W., H.S.W. and Q.S. performed the transport measurements. F.D., X.Z., X.C. and X.Z. performed the DFT calculations. C.C. and T.Y. performed the Raman measurements. H.W., X.X., S.T., F.D., C.C. and T.Y. analyzed the experimental data and designed the figures. H.W., S.T., C.C., T.Y., F.D., F.H., X.X. and M.J. co-wrote the manuscript and all authors contributed to critical discussions of the manuscript.

**Additional information:** Supplementary information accompanies this paper including: Description of graphene growth on *h*-BN, details of AFM measurements, classification of graphene domains grown on the surface of *h*-BN, Raman spectroscopy, device fabrication and transport characterization. This material is available free of charge via the Internet at http://.

**Competing financial interests**

The authors declare no competing financial interests.

# Supplementary Information

**Supplementary Figures:**

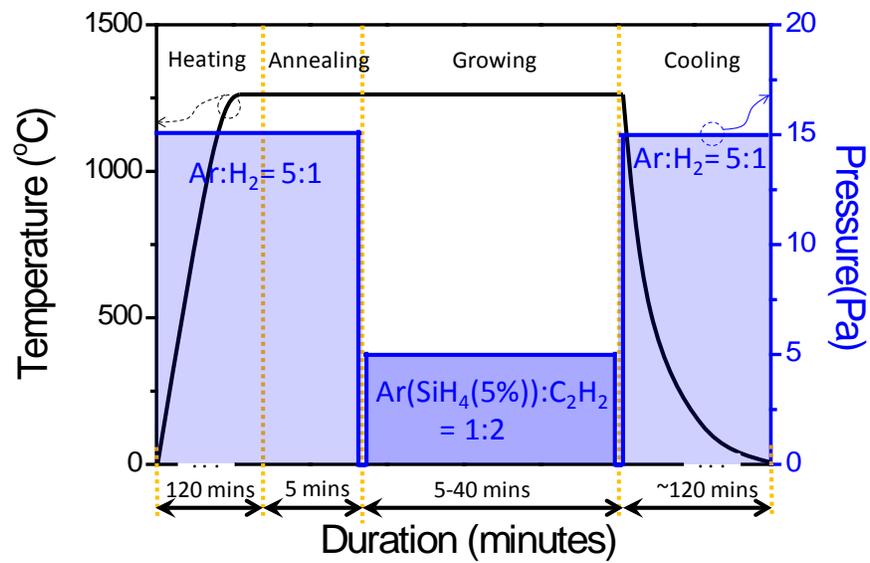

**Supplementary Figure 1 | Schematic diagram of a typical process for graphene growth on *h*-BN by GCA-CVD.**

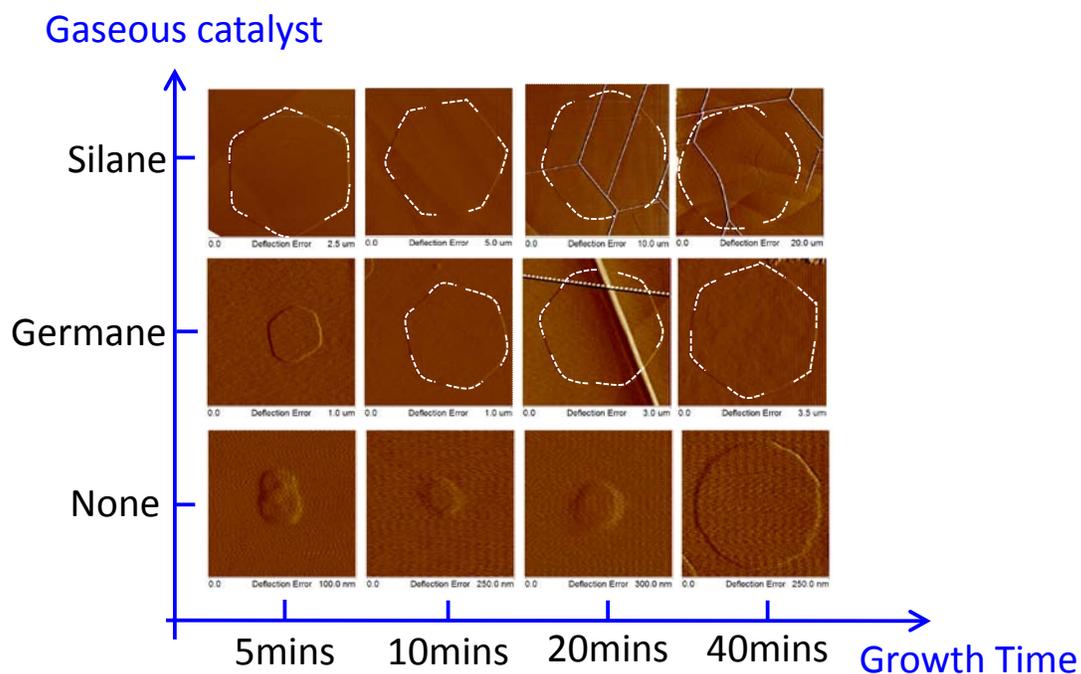

**Supplementary Figure 2 | Determining the growth rate of single crystalline graphene domain.** The size of single crystalline graphene is plotted as the function of the growth time in the presence of different gaseous catalysts. The dashed lines frame the shape of the graphene domains. All sizes of graphene domain are represented by the diagonal length of hexagonal graphene crystal. The growth temperature was kept at 1280°C.

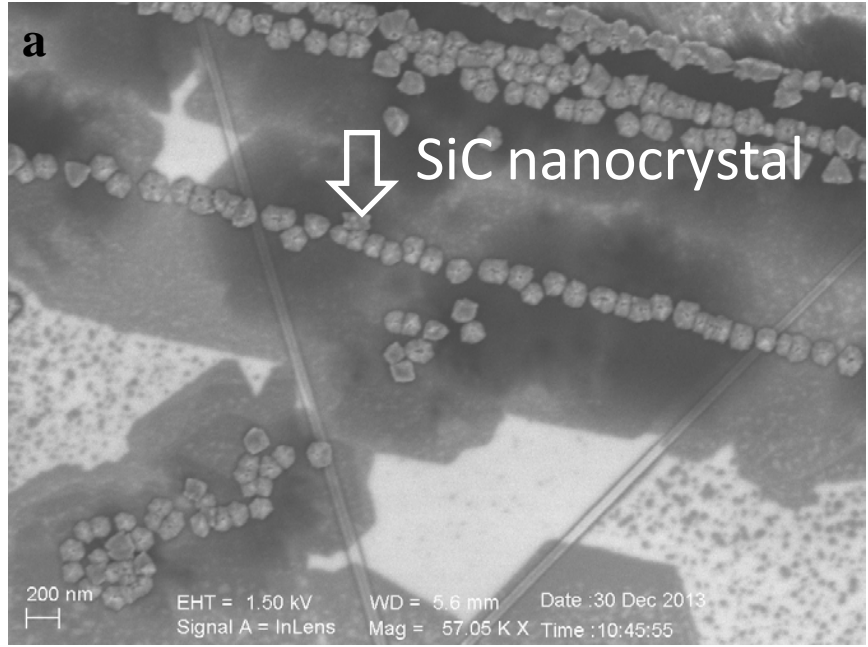

**Supplementary Figure 3 | SiC formation in high silane flow rate.** Characterization of the graphene materials grown for 10 minutes in the condition No.9 of Supplementary Table 1. (a) The image of scanning electron microscopy (SEM) on *h*-BN and (b) Raman spectrum of the graphene in *h*-BN. The SEM image shows that there are nano-crystals on *h*-BN, and the Raman characterization indicates the nano-crystals are mostly made from SiC.[1]

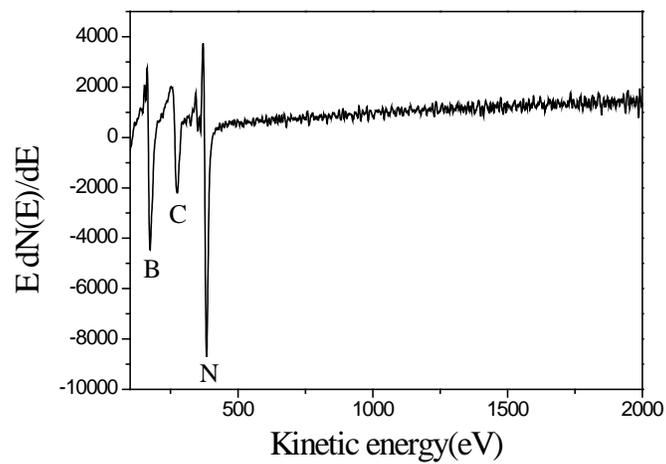

**Supplementary Figure 4 | Auger spectrum of graphene/*h*-BN (in an area of 10×10μm) in derivative mode plotted as a function of energy.** Different peaks for B, C and N are apparent. The survey spectrum shows the presence of boron, carbon and nitrogen. No obvious signal of Si/Ge is detected within the detection limit of the AES. These results indicate that the graphene grown on *h*-BN is almost Si/Ge-free.

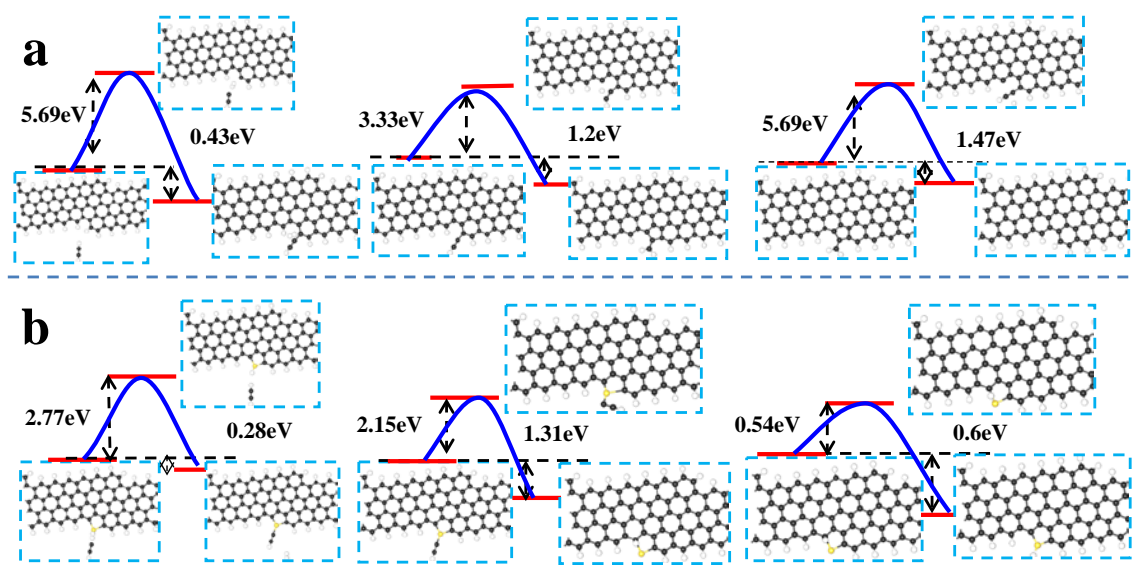

**Supplementary Figure 5 | The molecular dynamic simulation of the growth behaviors at the hydrogen terminated zigzag edge of graphene domains.** The repeatable cycles of incorporating a $C_2H_2$ molecule onto (a) H-terminated and (b) H- and Si- terminated graphene edges. Black, white and yellow balls represent carbon, hydrogen and silicon atom, separately.

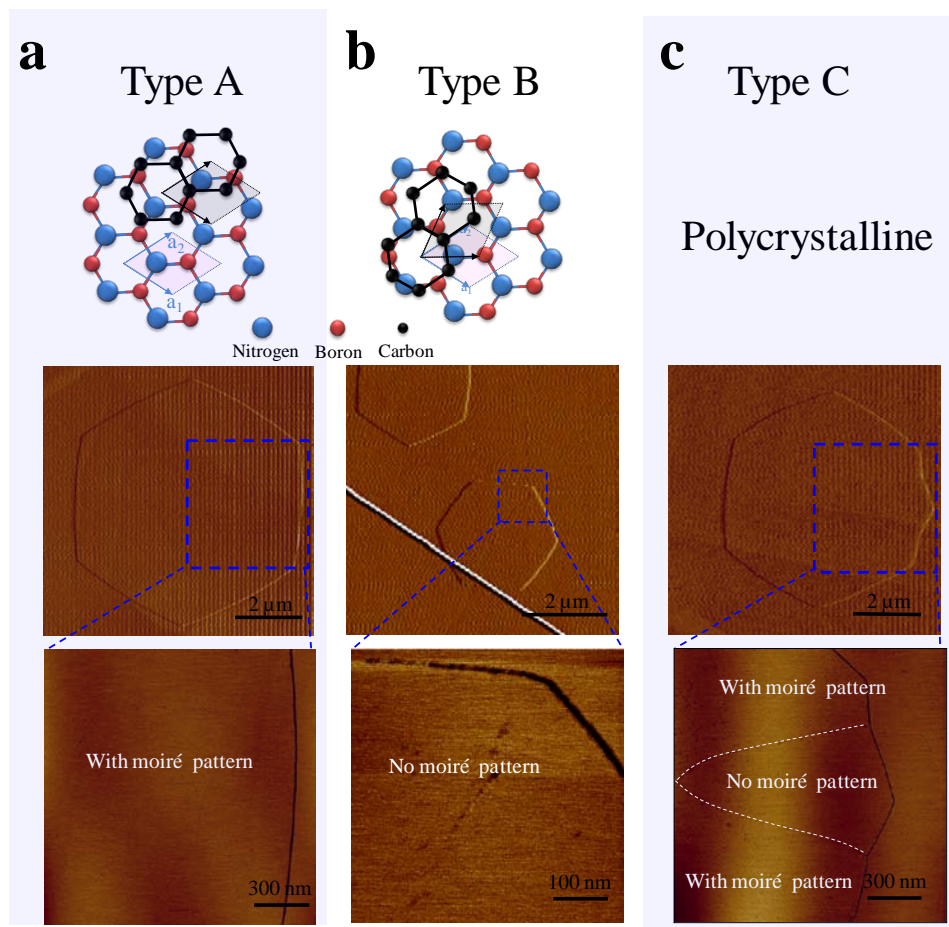

**Supplementary Figure 6 | Classification of graphene domains grown on the surface of *h*-BN.** The topography of the "different type" graphene domains on *h*-BN surface were measured. (a) Type "A" domain is precisely aligned with that of the underlying *h*-BN, and shows moiré pattern with the periodicity of about 13.9 nm; (b) Type "B" domain shows regular hexagonal shape but the graphene lattice is rotated about 30° relative to the underlying *h*-BN lattice; (c) Type "C" shows typical polycrystalline structure with moiré pattern detectable on some sub-domains but not on the others.

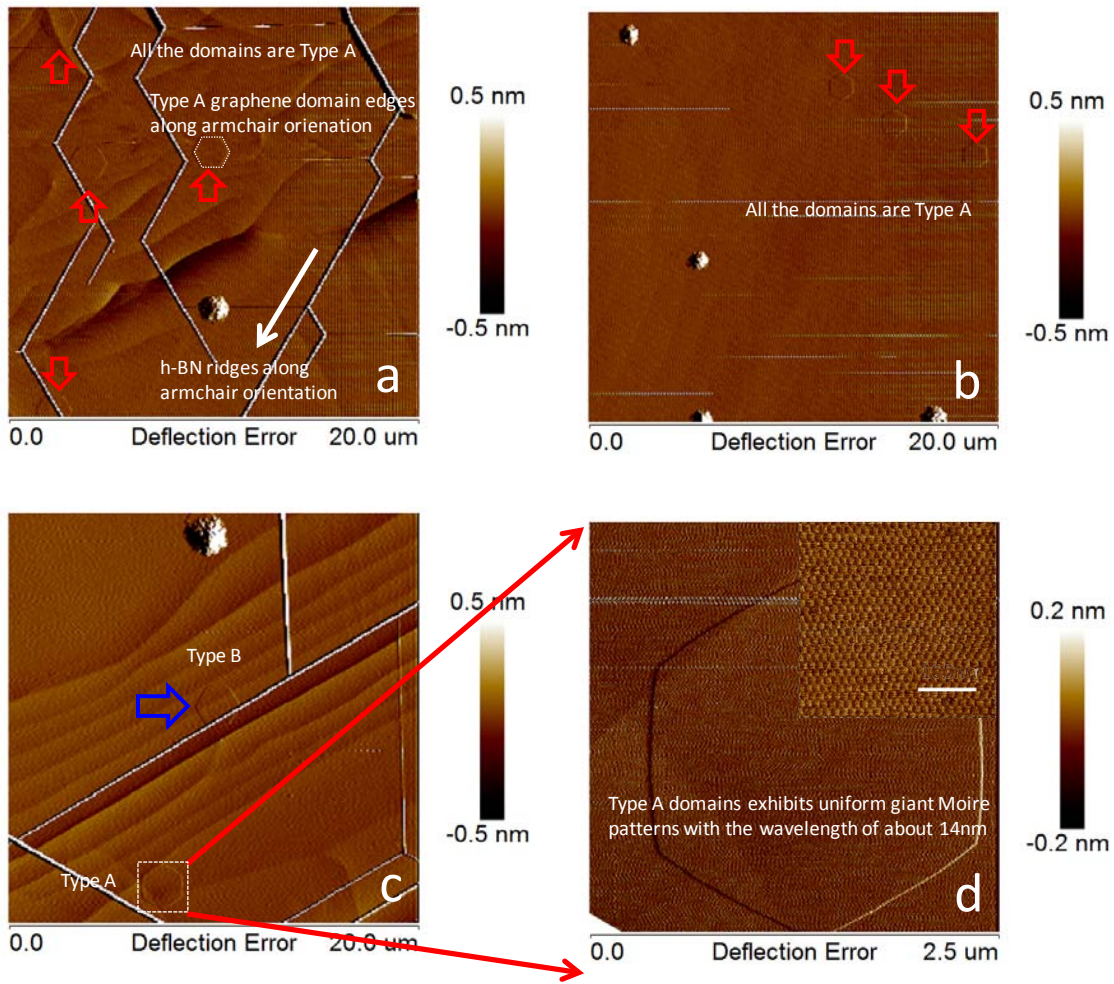

**Supplementary Figure 7 | Typical AFM images for graphene domains**. (a) & (b) typical AFM images for the "Type" survey, the domains are all "Type A" single crystalline graphene domain. The ridges are always used for the determination of the domain type; (c) The AFM image showing a "Type A" domain and a "Type B", where "Type A" exhibits uniform giant moiré patterns with a periodicity of 13.9 nm can be clearly seen (d).

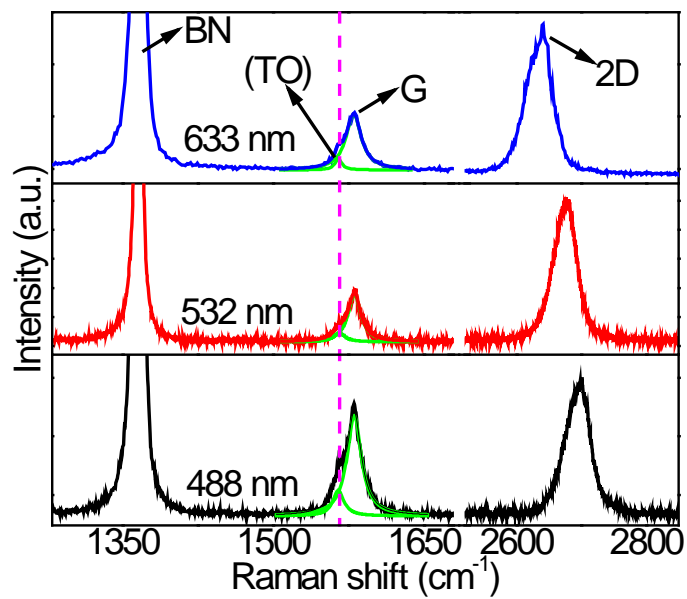

**Supplementary Figure 8 | Exciting energy dependence of the Raman spectrum of precisely aligned graphene.** Raman spectra were obtained with a WITec micro-Raman instrument possessing excitation laser lines of 488/532/633 nm. An objective lens of 100× magnification and a 0.95 numerical aperture (NA) was used, producing a laser spot that was ~0.5 µm in diameter. The laser power was kept less than 1 mW on the sample surface to avoid laser-induced heating. The excitation-laser-energy-dependent-Raman spectroscopy was used to support the conclusion that the shoulder peak located at 1565 cm$^{-1}$ in precisely aligned graphene domain is a TO phonon originated from the inter-valley umklapp scattering activated by graphene/$h$-BN super-lattice.

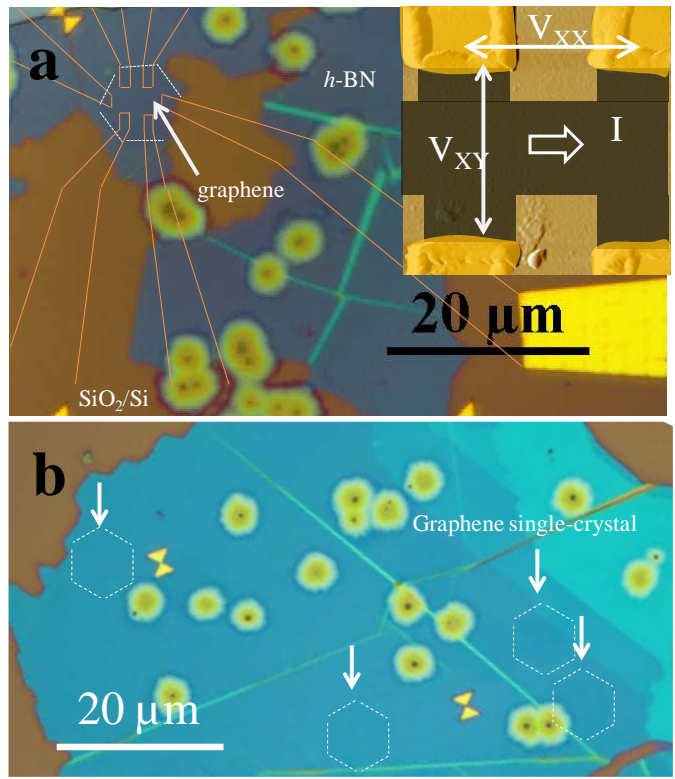

**Supplementary Figure 9 | Images of graphene domain and device design.** (a) Device design for a single crystal graphene domain grown on *h*-BN obtained by GVC-CVD, the inset shows a false-colored AFM image of a device on *h*-BN fabricated from the graphene single crystal grown on *h*-BN; (b) Optical image of a *h*-BN flake, the dashed white line marks the edges of single crystal graphene domains on *h*-BN.

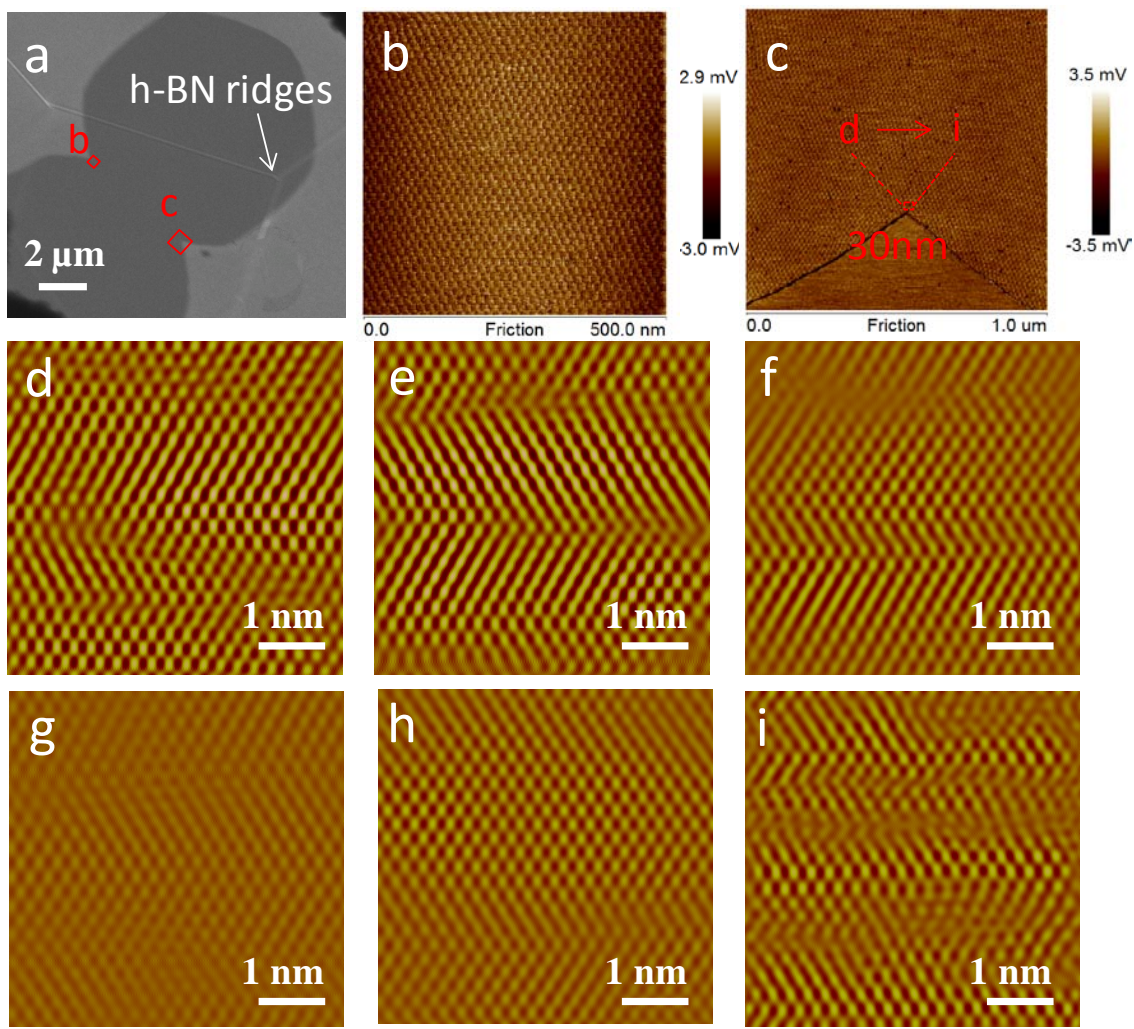

**Supplementary Figure** 10 | **Investigation on domain boundary where adjacent domains are coalescing.** (a) A SEM image of adjacent graphene domains on *h*-BN. Red boxes indicates the location where adjacent domains are merging. (b) and (c) AFM images of the boundary area within the red boxes of panel (a). The AFM images shows uniform moiré patterns distribution on the graphene domain. The moiré patterns exhibit good continuity. (d), (e), (f), (g), (h) and (i) are atomic scans (5 × 5nm$^2$) taken from red box (30 × 5 nm$^2$) in panel (c) (spatially continuous from left to right). These atomic images show the evolution of graphene domains to uniform monolayer during the CVD growth. By careful examination of atomic resolution AFM, no obvious domain boundary was found. These AFM images clearly show adjacent graphene domains have same lattice orientation and coalesce without grain boundary defects. It is confirmed that the adjacent domains merged at the boundary seamlessly.

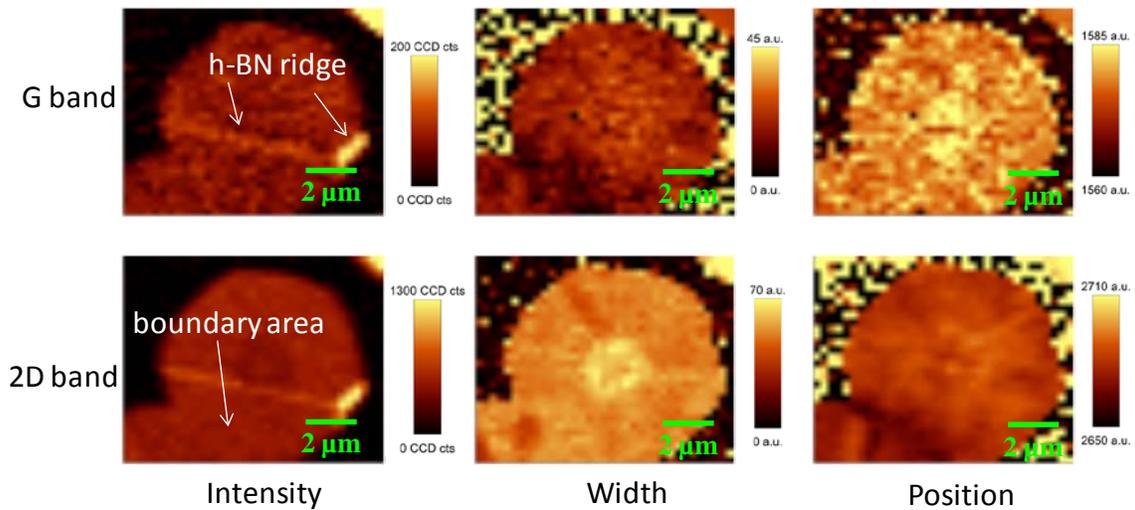

**Supplementary Figure 11 |** Raman G-band and 2D-band high-resolution-mapping of the grain boundary between two precisely aligned graphene domains. The high resolution Raman images show no obvious sign of grain boundary defects in the merged area. It is found that there is a very small red shift of 2D position in the merging area between domains. Other area is very uniform. The small shift of 2D peak (corresponding to < 0.1% strain) may attribute to the small stain introduced during the coalescing of graphene domains.

**Supplementary Tables:**

|  | Ar with 5%SiH$_4$ (sccm) | C$_2$H$_2$ (sccm) | Pressure (Pa) | Temperature (°C) | Duration (mins) | Graphene Diameter (μm) | Growth rate (μm/min) | Any SiC observed? |
|---|---|---|---|---|---|---|---|---|
| No.1 | 2 | 4 | 5 | 900 | 60 | 2.5 | 0.042 | No |
| No.2 | 4 | 4 | 5 | 900 | 60 | 2.6 | 0.043 | No |
| No.3 | 8 | 4 | 5 | 900 | 60 | 2.6 | 0.043 | No |
| No.4 | 16 | 4 | 5 | 900 | 60 | 2.8 | 0.047 | No |
| No.5 | 2 | 4 | 5 | 1280 | 20 | 4 | 0.2 | No |
| No.6 | 4 | 4 | 5 | 1280 | 20 | 6.8 | 0.34 | No |
| No.7 | 8 | 4 | 5 | 1280 | 20 | 7.5 | 0.375 | No |
| No.8 | 16 | 4 | 5 | 1280 | 20 | 8.2 | 0.41 | No |
| No.9 | 32 | 4 | 5 | 1280 | 20 | 8 | 0.4 | Yes |
| No.10 | 2 | 4 | 5 | 1350 | 10 | 10 | 1 | No |
| No.11 | 4 | 4 | 5 | 1350 | 10 | 9.8 | 0.98 | No |
| No.12 | 8 | 4 | 5 | 1350 | 10 | 9.5 | 0.95 | No |
| No.13 | 16 | 4 | 5 | 1350 | 10 | 9.1 | 0.91 | Yes |

**Supplementary Table 1 | Summary of CVD conditions evaluated for graphene growth on *h*-BN using gaseous catalyst.**

|  | G band | | (TO) band | | 2D band | |
| --- | --- | --- | --- | --- | --- | --- |
| Excitation Laser energy | Position (cm$^{-1}$) | FWHM (cm$^{-1}$) | Position (cm$^{-1}$) | FWHM (cm$^{-1}$) | Position (cm$^{-1}$) | FWHM (cm$^{-1}$) |
| 633 nm | 1580 | 18 | 1565 | 12 | 2635.6 | 49 |
| 532 nm | 1580.9 | 15 | 1565.8 | 13.7 | 2675.3 | 40 |
| 488 nm | 1580.6 | 17 | 1565 | 13 | 2695 | 44 |

**Supplementary Table 2 | Summary of peak position and full width at half maximum (FWHM) of the Raman peaks shown in Supplementary Fig. 8.**

| Year | Temp. | Mobility (cm$^2$/Vs) | Substrate | Geometry | Method | Growth rate | comments | Ref. |
|---|---|---|---|---|---|---|---|---|
| 2008 | 240K 5K | 120,000 170,000 | None | Hall and FET | Mechanically exfoliated from bulk graphite | N.A. | Freestanding graphene | 2 |
| 2009 | 1.6K | 3,750 | SiO$_2$ | FET | CVD on Cu | N.A. | Etching and Transfer needed | 3 |
| 2009 | Room Temp. (RT) | 4,050 | SiO$_2$ and Al$_2$O$_3$ | dual-gated FETs | Thermal CVD on copper | 3~5µm/min | Etching and Transfer needed | 4 |
| 2010 | RT | 800-16,000 | SiO$_2$ | FETs | Thermal CVD on copper | 3~10 µm/min | Etching and transfer needed | 5 |
| 2010 | 2K | 25,000 25,000-140,000 | h-BN | Hall mobility FET | Mechanically exfoliated from bulk graphite | N.A. | Transfer needed | 6 |
| 2011 | RT | 4,000 | SiO$_2$ | FETs | Thermal CVD | 0.1~4 µm/min | Etching and Transfer needed | 7 |
| 2011 | RT 2K | 3,000 10,500 | Al$_2$O$_3$ | Hall mobility | Thermal CVD | N.A. | Transfer free | 8 |
| 2011 | RT | 1800 | SiC | FET | Epitaxy and CVD | N.A. | Transfer free | 9 |
| 2011 | 4.2K | 9,200–28,800 4000–5400 | h-BN SiO$_2$ | Hall bar | CVD on Cu | 3~10µm/min | Etching and Transfer needed | 10 |
| 2011 | RT | <1000~10,000 | SiO$_2$ | FET | Thermal CVD on Cu | 1µm/min | Etching and transfer needed | 11 |
| 2012 | RT | 1,000–2,500 | SiO$_2$ | FET | Thermal CVD on Cu | 3.15µm/min | Etching and transfer needed | 12 |
| 2012 | RT | 7,100 | SiO$_2$ | FET | Thermal CVD on Pt | 4.17µm/min | Etching and transfer needed | 13 |
| 2012 | 1.6K | 27,200~44,900 | h-BN | Hall bar | Thermal CVD on Cu | 0.67~2.1µn/min | Etching and transfer need | 14 |
| 2012 | RT | 277± 91(e) 227 ± 66(h) | Al$_2$O$_3$ | FET | Thermal CVD | 2.78nm/min | Transfer free | 15 |
| 2012 | RT | 2,000 | Al$_2$O$_3$ | Hall mobility | Thermal CVD | 50nm/min | Transfer free | 16 |
| 2013 | RT | 1,300-5,650 | Si$_3$N$_4$/SiO$_2$/Si | FET | Oxygen-Aided CVD | 2.5nm/min | Transfer free and polycrystalline | 17 |
| 2013 | 1.5K | 5,000 | h-BN | FET | Plasma-enhanced CVD | 0.65nm/min | Transfer free and well aligned | 18 |
| 2013 | Low T | 20,000-80,000 | h-BN | FET and Hall | Mechanically exfoliated from bulk graphite | N.A. | Transfer needed | 19 |
| 2013 | Low T | 10,000~100,000 | h-BN | FET and Hall | Mechanically exfoliated from bulk graphite | N.A. | Transfer needed | 20 |
| 2014 | RT | 531 | SiO$_2$ | FET | CVD on | 1.66nm/min | Transfer free | 21 |
| 2014 (this work) | RT | ~17,000 19,000(e)-23,000(h) | h-BN | FET and Hall | Gaseous Catalyst Assisted-CVD | 1µm/min | Transfer free and well aligned | this work |

**Supplementary Table 3 | Survey on mobility of charge carriers in graphene reported in literatures.** The properties of CVD graphene samples, especially the electronic performance, are highly dependent on factors such as synthesis and processing techniques, and adjacent materials. The survey in carrier mobility shown here indicates that the mobility prefers the flat *h*-BN with inert surface.

# Supplementary Notes

**Supplementary Note 1 | the effect of solid catalyst**

To qualitatively understand the role of silicon (germanium) atom in the growth of graphene, solid silicon, solid germanium and their alloy were placed near $h$-BN flakes, respectively. The Ge/Si alloy consists of 30% Si in molar ratio. We found that the presence of solid silicon/germanium/their alloy can effectively improve growth rate of graphene when heating them to 1280 ℃ or higher. It indicates that silicon/germanium vapor plays a role of catalyst in graphene growth on $h$-BN. The results illuminate us that the saline/germane may be effective catalysts for graphene growth on $h$-BN.

**Supplementary Note 2 | Key parameters for the graphene growth**

The graphene growth was influenced mainly by the gaseous catalyst and growth temperature:

1. Argon/silane mixture flow rate: a $C_2H_2$ flow and a mixture of silane/argon (the mole ratio of silane to argon was 5%) were introduced into the system for the graphene growth. High flow rate of the argon/silane mixture resulted in a rapid growth rate. Some growth results with different argon/silane mixture flow rate are summarized in Supplementary Table 1.

2. Growth temperature: Temperature plays an important role in decomposition of hydrocarbons and activates the catalyst. An increase in the growth temperature provides the possibility for increasing the growth rate. Some growth results with different growth temperature are summarized in Supplementary Table 1. The quality of the graphene grown under different temperature shows little difference via Raman measurement.

3. Too much silane could lead to the formation of SiC as shown in Supplementary Fig. 3.

**Supplementary Note 3 | $C_2H_2$ as a precursor.**

It is worthy to note that the above picture hold valid only when $C_2H_2$ is used as a precursor. If we use $CH_4$ as a carbon precursor, the atmospheric composition in the reaction chamber will be very different. We experimentally verified that the catalytic effect is not obvious, if using $CH_4$. The catalytic effects due to silicon or germanium atoms are further validated by heating solid silicon or Si/Ge alloys to different temperatures at different pressures, thus producing different vapor pressures of Si and Ge atoms.

## Supplementary Discussions

**Simulation of the growth behaviors at zigzag edges.** To get a more detailed understanding on the growth mechanism at zigzag edges, we performed a DFT calculation and the results are shown in Supplementary Fig. 5. We assumed that the growth frontier is passivated by hydrogen atoms, and the $C_2H_2$ molecule is the carbon feedstock. In the non-catalyst situation (Supplementary Fig. 5a), the $C_2H_2$ molecules need three steps to incorporate to the graphene edge to form a new carbon ring. The energy barriers for the three steps are 5.69eV, 3.33eV and 5.69eV, respectively. After the new carbon ring is formed, the growth frontier steps forward and form a same structure as that before the $C_2H_2$ molecule integrated in the carbon ring. In the silicon catalyst situation (Supplementary Fig. 5b), we start with one silicon atom being absorbed on the growth frontier, in the growth circle such silicon atom serves as a bridge between carbon dimer and graphene, which lowers the energy needed for dehydrogenation of $C_2H_2$ and formation of carbon-carbon bonds to 2.77eV, 2.15eV and 0.5eV. The much lowered energy barriers account for the greatly enhanced growth rate.

**Classification of graphene domains grown on the surface of *h*-BN.** Actually, we scanned the surface of *h*-BN flakes and measured the topography of the "different type" graphene domains on *h*-BN surface by AFM. We measured about 200 pieces of *h*-BN flakes for the statistics in Fig. 3. In total, about 1600 graphene domains are measured. There are some advices in such an investigation. 1) The edges orientation of graphene domain, moiré patterns and the atomic resolution images of graphene on *h*-BN help to determine the type of graphene domains. The details about the investigating methods are introduced in our earlier publication (S. Tang et al., Precisely aligned graphene grown on hexagonal boron nitride by catalyst free chemical vapor deposition. Sci. Rep. 3, 2666 (2013)). 2) There are always ridges formed on the surface of *h*-BN. The *h*-BN ridges are caused by the different thermal expansion coefficients between *h*-BN and underling $SiO_2$ surface. We found that almost all the ridges are precisely along the armchair direction of h-BN. As most of the graphene domains are of edges in parallel with armchair direction, one could make use of the ridges to know the lattice orientation of h-BN. It really can save much time. 3) Graphene single crystalline domains always exhibit a hexagonal shape. The domains show a giant superlattice with wavelength of about 14 nm if the domains are precisely aligned with the h-BN. 4) the superlattices are also known as "moiré patterns", and high quality graphene domains have very uniform moiré patterns.

Although the investigations are actually tedious and time-consuming, it is worthy to do such an investigation to know the uniformity of the graphene domain. Supplementary Fig. 7 gives more examples for the explanation of the type survey.

**Transport measurement.** Electronic transport measurements in a Hall bar configuration were also carried out to characterize the graphene single crystal grown on h-BN. The gate voltage ($V_g$) dependence of the longitudinal resistance ($R_{xx}$) at different temperature is plotted in Fig. 5a. The main peak in $R_{xx}$ at $V_g = -5$ V, represents graphene's main neutrality point. Two satellite peaks symmetrically appear on both sides of the main neutrality point. As the temperature decreases, these two satellite peaks, become more obvious. The satellite peak on the hole side appears much stronger than that on the electron side. The satellite peaks in transport properties are believed to be related to the spectral reconstruction in graphene brought to contact with h-BN. The superlattice potential induced by the h-BN results in the appearance of secondary Dirac points (SDP) in graphene's energy dispersion. They can be observed as the satellite peaks in the transport measurement after the Fermi energy is tuned to reach the reconstructed part of the spectrum. The resistances ($R$) at the Dirac Point (DP) and satellite peak at the hole branch as a function of $T$ are plotted in the inset of Fig.5a, both of them exhibit very weak temperature dependence, the results are different from earlier reports. One possible reason is that the commensurate state is suppressed.[24]

As the satellite resistance peaks in the transport data originates from the moiré pattern, the wavelength of moiré pattern can be estimated by measuring the relative position of the satellite resistance peaks. As the energy separation between the DP and the secondary

Dirac point (SDP) is $E_{SDP} = \frac{hv_F}{\sqrt{3}\lambda}$, where $\lambda$ is superlattice period, $v_F$ is graphene's Fermi velocity and $h$ is Planck's constant. The Fermi level can be tuned by applying the external gate voltage: $E_F = \frac{h}{2\pi}v_F\sqrt{\pi C_g(V_g - V_{NP})/e}$, where "$e$" represents elementary charge, the effective capacitance $C_g$ can be estimated at about 10.5 nF·cm$^{-2}$. Thus, we can roughly estimate the wavelength of the morie pattern from the equation $\lambda = 2\sqrt{\frac{\pi e}{3C_g(V_g - V_{NP})}}$, where the $V_g$ represents gate voltage, $V_{NP}$ represents gate voltage at neutral point. Given that $V_g - V_{NP} = 33$ V and $C_g \approx 10.5$ nF/μm$^2$, we can get $\lambda \approx 14.1$ nm. The wavelength of the morie pattern derived from the gate voltage at which the secondary Dirac point occur matches well with our results obtained from AFM measurement.[25,26]. From Fig. 5a, the electrical field mobility at 300 K is about 17,000 cm$^2$V$^{-1}$s$^{-1}$, which is extracted from a density independent mobility model $R_{total} = \frac{L/W}{\mu\sqrt{(n_0 e)^2 + (V_g - V_{dirac})^2 C_g^2}}$, where $n_0$ is the residual carrier density induced by charge impurities.[27] $L$ and $W$ represent channel length and width, respectively. Both of them equals to 1 μm in this device. Here, the dielectric constant of $h$-BN is comparable with that of SiO$_2$. Magnetotransport for the graphene/$h$-BN heterostructure is also measured at T≈ 300 K. Longitudinal resistance ($R_{xx}$) and Hall resistance ($R_{xy}$) as a function of gate voltage taken in a magnetic field of $B = 9$ T are shown in Fig. 5b. The extracted Hall mobility $\mu = \frac{R_{xy} \cdot L}{R_{xx} \cdot W} \cdot \frac{1}{B}$ is about 19,000 cm$^2$V$^{-1}$s$^{-1}$ for the holes and ~23,000 cm$^2$/V·s for electrons.

The color plots of the $R_{xx}$ and $R_{xy}$ as a function of both gate voltage and magnetic field are shown in Fig.5c and 5d, respectively. The standard quantum Hall effect (QHE) for graphene, is observed with valleys in $R_{xx}$ (Fig.5c) and plateau in $R_{xy}$ (Fig. 5d) at filling factors $\nu = \pm 4(n+1/2) = \pm 2, \pm 6, \pm 10$ and $\pm 14,...$ where $n = 0, 1...$ is the LL index. The features are the characteristic of the gapless Dirac spectrum of graphene. Similar features fan out from secondary Dirac point in the precisely aligned graphene on $h$-BN and the Dirac Fermion physics near the main Dirac point is unperturbed. It is noted that the resistance peak of the SDP on the hole doping regime broadens with the increase of magnetic field. Over all, the half integer QHE observed in the precisely aligned graphene on *h*-BN indicates the high quality of precisely aligned graphene grown on *h*-BN.

## Supplementary Reference